\documentclass[sigconf]{acmart}

\AtBeginDocument{%
  \providecommand\BibTeX{{%
    \normalfont B\kern-0.5em{\scshape i\kern-0.25em b}\kern-0.8em\TeX}}}

\usepackage{booktabs} 
\usepackage{xcolor}
\newcommand{\cmmnt}[1]{}
\usepackage{todonotes}
\setcopyright{none}

\copyrightyear{2020}
\acmYear{2020}

\acmConference[This is an author version submitted to ACM Hypertext 2020]{Submitted to ACM Hypertext '20}{}{}
\acmDOI{10.1145/3372923.3404781}
\acmISBN{978-1-4503-7098-1/20/07}
\acmBooktitle{Submitted to ACM Hypertext '20}
\begin{document}


\title{Off-line vs. On-line Evaluation of Recommender Systems in Small E-commerce}

\author{Ladislav Peska}
\affiliation{
  \institution{Department of Software Engineering, \\Faculty of Mathematics and Physics, Charles University}
  \streetaddress{Malostransk\'e n\'am. 25, Prague, 11800, Czech Republic}   
}
\email{peska@ksi.mff.cuni.cz}

\author{Peter Vojtas}
\affiliation{
  \institution{Department of Software Engineering, \\Faculty of Mathematics and Physics, Charles University}
  \streetaddress{Malostransk\'e n\'am. 25, Prague, 11800, Czech Republic}   
}
\email{vojtas@ksi.mff.cuni.cz}


\begin{abstract}

In this paper, we present our work towards comparing on-line and off-line evaluation metrics in the context of small e-commerce recommender systems. Recommending on small e-commerce enterprises is rather challenging due to the lower volume of interactions and low user loyalty, rarely extending beyond a single session. On the other hand, we usually have to deal with lower volumes of objects, which are easier to discover by users through various browsing/searching GUIs.

The main goal of this paper is to determine applicability of off-line evaluation metrics in learning true usability of recommender systems (evaluated on-line in A/B testing). In total 800 variants of recommending algorithms were evaluated off-line w.r.t. 18 metrics covering rating-based, ranking-based, novelty and diversity evaluation. The off-line results were afterwards compared with on-line evaluation of 12 selected recommender variants and based on the results, we tried to learn and utilize an off-line to on-line results prediction model.

Off-line results shown a great variance in performance w.r.t. different metrics with the Pareto front covering 64\% of the approaches. Furthermore, we observed that on-line results are considerably affected by the seniority of users. On-line metrics correlates positively with ranking-based metrics (AUC, MRR, nDCG) for novice users, while too high values of  novelty had a negative impact on the on-line results for them.
\end{abstract}

\maketitle

\section{Introduction}
Recommender systems (RS) belong to the class of automated content-processing tools, aiming to provide users with unknown, surprising, yet relevant objects without the necessity of explicitly query for them. The core of recommender systems are machine learning algorithms applied on the matrix of user to object preferences. As such, recommender systems are highly studied research topic as well as extensively used in real-world applications.

However, throughout the decades of recommender systems research, there was a discrepancy between industry and academia in evaluation of proposed recommending models. While academic researchers often focused on off-line evaluation scenarios based on recorded past data, industry practitioners value more the results of on-line experiments on live systems, e.g., via A/B testing. While off-line evaluation is easier to conduct, repeatable, fast and can incorporate arbitrary many recommending models, it is often argued that it does not reflect well the true utility of recommender systems as seen in on-line experiments \cite{rev-OfflineAB}.
On-line evaluation is able to naturally incorporate current context, tasks or search needs of the user, appropriateness of recommendations' presentation as well as causality of user behavior. On the other hand, A/B testing on live systems is time consuming, the necessary time scales linearly with the volume of evaluated approaches and it can even harm retailer's reputation if bad recommendations are shown to users.

\subsection{Bridging the Off-line vs. On-line Gap}
A wide range of approaches aimed to bridge the gap between industry and academia.
Jannach and Adomavicius \cite{rev-JanAdom-userStudies} argue for recommendations with a purpose, i.e., after a certain level of RS's maturity. In particular, after the numerical estimators of user's preference are established, authors suggest to step back and revisit some of the foundational aspects of RS. Authors aimed to reconsider the variety of purposes, for which recommender systems are already used today in a more systematic manner and  proposed a framework which should cover both consumer's/provider's viewpoint and strategic/operational perspective. 
One way to approach this goal are user studies via questionnaires (e.g. \cite{rev-userStudQuestionaire}) or more involved frameworks, e.g. \cite{rev-ZenoUserExperience}. Still, the main problem remains: we may lack the participants, whose motivation, information needs and behavior would be similar to real-world users. 

Another approach to treat the off-line/on-line phenomenon comes from considerations about relevance of statistical learning in understanding causation, confounding, missing (not at random - MNAR) data  (see e.g., \cite{rev-MissingData}). 
A starting point of these approaches is the observation that implicit feedback (despite many advantages) has inherent biases and these are key obstacles to its effective usage. For example, position bias in search rankings strongly influences how many clicks a result receives, so that directly using click-through data as a training signal in Learning-to-Rank (LTR) methods yields sub-optimal results \cite{4292009}. To overcome the bias problem, Joachims et al. \cite{rev-JoachSwa-Unbiased} presented a counterfactual inference framework that provides the theoretical basis for unbiased LTR via Empirical Risk Minimization despite the biased data. 
Also Gilotte et al. \cite{rev-OfflineAB} utilized de-biased off-line methods to estimate the potential uplift of the on-line performance. 
     Authors proposed a new counterfactual estimator and evaluated it on a proprietary dataset of 39 past A/B tests, containing several hundreds of billions of recommendations in total.

A recent contribution to academia{-}industry discussion was the 2017 Recommender Systems Challenge \cite{rev-intro-2017RecSysChallenge}, focused to the problem of job recommendations\footnote{http://www.recsyschallenge.com/2017/}. In the first phase, participants evolved their models on off-line data. Afterwards, invited participants were tasked to provide and evaluate recommendations on-line. Most of the teams managed to preserve their off-line performance also during the on-line phase. Quite surprising was the fact that traditional methods and metrics to estimate the users' preferences  for unknown items (of course, tuned to specifics of the task) worked best. The winning team combined content and neighbor{-}based models with feature extraction, balanced sampling and minimizing a tricky classification objective \cite{rev-VitezRSChallVolkovs}. 

\subsection{Recommender Systems in Small E-commerce}
Previously mentioned approaches were mostly user-centric. However, in the RecSys Challenge 2017 \cite{rev-intro-2017RecSysChallenge}, we could observe the success of item-based methods. The main cause was probably the cold start problem, which is prevalent also in small e-commerce enterprises. 
Kaminskas et al. \cite{rev-Kam-SME-JDataSem} observed that the small amount of returning customers makes traditional user-centric personalization techniques inapplicable and designed an item-centric product recommendation strategy. Authors deployed the proposed solution on two retailers' websites and evaluated it in both on-line and off-line settings.  
Jannach et al. \cite{rev-JanachShortTerm-Conference_RecSys_2015_st} considered the problem of recommending to users with short{-}term shopping goals. Authors observed the necessity of item{-}based approaches but also importance of algorithms usually used for long{-}term preferences. 

Peska and Vojtas \cite{rev-PV-JDataSem} proposed the usage of implicit preferences relations on the problem of recommending for small e-commerce enterprises with short-term user's goals. Their work is based on an complex observation of users' behavior up to the level of noticeability of individual objects on the category pages.

In general, providing recommendation service on small e-commerce enterprises brings several specific challenges and opportunities, which changes some recommending paradigms applied, e.g., in large-scale multimedia enterprises. Let us briefly list the key challenges:
\begin{itemize}
  \item High competition has a negative impact on user loyalty. Typical sessions are very short, users quickly leave to other vendors, if their early experience is not satisfactory enough. Only a fraction of users ever returns.
  \item For those single-time visitors, it is not sensible to provide any unnecessary information (e.g., ratings, reviews, registration details). 
  \item Consumption rate is low, users often visit only a handful (0-5) of objects and rarely ever buys anything.
  \item Small e-commerce enterprises generally offer lower volume of objects (ranging usually from hundreds to tens of thousands instead of millions as in, e.g., Amazon).
  \item Objects often contain extensive textual description as well as a range of categorical attributes. Browsing and attribute search GUIs are present and widely used.
\end{itemize}
The first three mentioned factors contribute to the data sparsity problem and limit applicability of user-based collaborative filtering (CF). Although the total number of users may be relatively large (hundreds or thousands per day), the volume of visited objects per user is limited and the timespan between the first and last feedback is short.
The last two factors contribute towards objects' discoverability. This may seemingly decrease the necessity of recommender systems\footnote{Although objects are more discoverable and users do not depend on recommendations only, they are often not willing to spend too much time in the discovery process and recommendations may considerably shorten it.}, but also decreases the effect of missing not at random data \cite{rev-MNAR-p5-marlin} and therefore may contribute to the consistency of off-line and on-line evaluation. Also, in many product domains (including our travel agency test bed), it is uncommon to have any "well-known" items, such as blockbuster movies or popular songs. This further limits applicability of counterfactual approaches.

Despite mentioned obstacles, the potential benefit of recommender systems in small e-commerce enterprises is still considerable, e.g., "more-of-the-kind" and "related-to-purchased" recommendations are not easy to mimic with standard search/browsing GUI. 

\subsection{Off-line to On-line Predictions}
Garcin et al. \cite{RS2014_news} focused on news recommendations and observed a major difference between off-line and on-line accuracy evaluations. These differences went beyond a small numerical variance and had a determining impact on the ordering of best methods. Utilized metric (hit@top-3) is somewhat proprietary, but supported by the website design. In a follow-up study \cite{RS2015_news}, authors focused on additional off-line metrics including accuracy, diversity, coverage and serendipity metrics. Similarly as in \cite{RS2015_news}, we usually observed very high correlation scores for metrics from a single cluster (ranking-prediction, rating-prediction), but correlations between metrics from different clusters were much lower in our case. This work also inspired us to employ regularized linear regression models to predict on-line performance from off-line results.

Rossetti et al. \cite{RS16_user_study} focused on the MovieLens dataset and organized a user study aiming to compare off-line and on-line evaluation metrics. Authors specifically distinguished long-tail recommendations and recommendations of previously unknown items. Similarly as in \cite{RS2014_news}, same metrics were evaluated off-line and on-line. Authors showed that off-line evaluation induces similar ranking of algorithms, but with some exceptions. Also Beel et al. \cite{RS_libraries2015} focused on ranking accuracy metrics such as nDCG and MRR in a literature RS. Authors reported on some moderate correlations between CTR and these off-line metrics, but also mentioned several cases, where the prediction failed. 
Gruson et al. \cite{wsdm19_playlists} focused on the problem of candidates selection for on-line evaluation in Spotify playlists recommendations. Authors employed several approaches to de-bias the off-line evaluations based on importance sampling, where some approaches have seemingly good prediction results. However, authors did not compare these models with original "biased" feedback. Therefore, it is hard to asses the importance of feedback de-biasing. As the nature of small e-commerce domains seemingly reduces such feedback biases, we did not include such approaches in our current study, but we plan to explore them in the future work. Let us also note that none of \cite{RS_libraries2015, RS16_user_study, wsdm19_playlists} considered other off-line metrics than some form of ranking accuracy.

Although the mentioned related studies (as well as our own work) share the general goal of observing and describing relations between off-line and on-line results, there is a determining difference in the considered application domains. This has an effect on both the choice of recommending algorithms, evaluation metrics as well as the generic study design. Specifically, we are not aware of any work considering the predictability of on-line results from off-line metrics in the context of small e-commerce enterprises. We also evaluated a wide range of off-line metrics beyond ranking accuracy and evaluated the effect of promoting diversity or novelty of recommendations. We would also like to note that mentioned papers only considered a single class of recommending algorithms, while in our work, we evaluated several diverse recommending algorithms.

\subsection{Main Contributions}
 The main scope of this paper is to contribute towards bridging the gap between industry and academia in the evaluation of recommender systems. We specifically focused on the domain of small e-commerce enterprises and within this scope, we aim on determining the usability of various off-line evaluation methods and their combinations in learning the relevance of recommendations w.r.t. on-line production settings. In total, 800 variants of recommender systems (3 base recommending algorithms combined with 9 user profile construction algorithms and various hyperparameter settings) were evaluated off-line w.r.t. 18 metrics covering rating-based, ranking-based, novelty and diversity metrics. The off-line results were afterwards compared with on-line evaluation of 12 selected algorithm's variants.

To sum up, main contributions of this paper are as follows:

\begin{itemize}
  \item We compared a wide range of off-line metrics against the actual on-line results w.r.t. click through rate (CTR) and visits after recommendation (VRR). 
  \item The observed results highly depend on users "seniority". While, the ranking-based metrics generally correlate with on-line results for less senior users, novelty and diversity gain importance for users with more visited objects. 
  \item We further evaluated several simple regression techniques aiming to predict on-line results based on the off-line metrics and achieved considerable predictability of CTR and VRR under leave-one-out cross-validation (LOOCV) scenario.  
  \item Based on the previous point, we may recommend \linebreak word2vec and some variants of cosine CB methods to be used on small e-commerce enterprises. 
  \item Datasets acquired during both off-line and on-line evaluation are available for future work.
\end{itemize}

\section{Materials and Methods}
\subsection{Dataset and Evaluation Domain}
As the choice of suitable recommending algorithms is data-dependent, let us first briefly describe the dataset and the domain, we used for evaluation.

Experiments described in this paper were conducted on a medium-sized Czech 
travel agency. The agency sells tours of various types to several dozens of countries. Each object (tour) is available in selected dates. Some tours (such as trips to major sport events) are one-time only events, others, e.g., seaside holidays or sightseeing tours are offered on a similar schedule with only minimal changes for several years. All tours contain a textual description accompanied with a range of content-based (CB) attributes, e.g., tour type, meal plan, type of accommodation, length of stay, prices, destination country/ies, points of interest etc.

The agency's website contains simple attribute and keyword search GUI as well as extensive browsing and sorting options. Recommendations are displayed on a main page,  browsed categories, search results and opened tours. However, due to the importance of other GUI elements, recommendations are usually placed below the initially visible content.

\subsection{Recommending Algorithms}
In accordance with Kaminskas et al. \cite{rev-Kam-SME-JDataSem}, we considered user-based recommending algorithms, e.g., matrix factorization models impractical for small e-commerce due to a high user fluctuation and short timespan between user's first and last visits. 

\subsubsection{\textbf{Item-to-item Recommending Models}}
We considered three recommending approaches corresponding with the three principal sources of data: object's CB attributes, their textual description and the history of users' visits (collaborative filtering). The information sources are mostly orthogonal, each focused on a different recommending paradigm. The expected output of recommendations based on CB attributes is to reveal similar objects to the ones in question. By utilizing the stream of user's visits, it is possible to uncover objects that are related, yet not necessarily similar. The expected outcome of textual-based approaches is also to provide similar objects, however the similarity may be hidden within the text, e.g., seaside tours with the same type of beach, both suitable for families, located in a small peaceful village, but in a different country.
For each type of source information, we proposed one recommending algorithm as follows:

\medskip
\noindent
--  Skip-gram \textbf{word2vec} model \cite{rev-word2vec} utilizes the stream of user's visits. Similarly as in \cite{DBLP:journals/corr/BarkanK16}, the sequence of visited objects is used instead of a sentence of words, however, we kept the original window size parameter in order to better model the stream of visits. The output of the algorithm is an embedding of a given size for each object, while similar embeddings denotes objects appearing in a similar context. In evaluation, embedding's size was select from \{32, 64, 128\} and context window size was selected from \{1, 3, 5\}.

--  \textbf{Doc2vec} model \cite{DBLP:journals/corr/LeM14} utilizes the textual description of objects. Doc2vec extends word2vec model by an additional attribute defining the source document (object) for each word in question. The model, in addition to the word embeddings calculates also embeddings of the document itself, therefore the output of the algorithm are embeddings of a given size for each object (document). Textual descriptions of objects were preprocessed by a 
stemmer \footnote{Language and link removed for the sake of anonymization
} and stop-words removal. In evaluation, embedding's size was select from \{32, 64, 128\} and window size from \{1, 3, 5\}.

-- Finally, we used \textbf{cosine similarity} on CB attributes. Nominal attributes were binarized, while numeric attributes were standardized before the similarity calculation. We evaluated two variants of the approach differing in whether to allow evaluating similarity on self\footnote{Otherwise, the similarity of an object to itself is zero by definition.}. In this way, we may promote/restrict recommendations of already visited objects, which belongs to some of the commonly used strategies.

Given a query of a single object, the base recommended list would be a list of top-k objects most similar to the query object (or its embeddings vector).

\subsubsection{\textbf{Using History of User's Visits}}
While the above described algorithms focus on modeling item-item relations, we may posses a longer record of visited objects for some users. Although many approaches focused on a last visited object only, e.g., \cite{rev-Kam-SME-JDataSem}, some approaches using the whole user session emerged recently \cite{DBLP:journals/corr/HidasiKBT15}.

Therefore, we proposed in total nine methods to process users' history and aggregate recommendations for individual objects. The variants are as follows:

\begin{itemize}
  \item Using \textbf{mean} of recommendations for all visited objects. 
  \item For each candidate object, use \textbf{max} of its similarity w.r.t. some of the visited object. 
  \item Using \textbf{last} visited object only. 
  \item Using weighted average of recommendations with linearly decreasing weights. In this case, only the \textbf{last-k} visited objects are considered, while its weight $w=1-(rank/k)$ linearly decreases for older visits. We evaluated results considering last 3, 5 and 10 objects.
  \item Using weighted average of recommendations with \textbf{temporal} weights. This variant is the same as the previous one, except that the weights of objects are calculated based on the timespan between the current date and the date of visit: $w = 1/(log(timespan.days) + \epsilon)$. We evaluated results considering last 3, 5 and 10 objects as well as a full user profile.
\end{itemize}

While the first two approaches considered uniform importance of the visited objects, others rely on some variations of \textit{"the newer the better"} heuristic. Using history of the user instead of the last item only is one of the extensions of our work compared to \cite{rev-Kam-SME-JDataSem}.

\subsubsection{\textbf{Novelty and Diversity Enhancements}}
\label{sec:novelty_enhancements}
The per\-for\-man\-ce of recommenders may also depend on a lot of subjective, user-perceived criteria, as introduced in \cite{Ricci:2011}, such as \textit{novelty} or \textit{diversity} of recommended items. Therefore, in the off-line evaluation (Section \ref{sec:off_line_eval}, we evaluated one type of diversity metric (intra-list diversity \cite{DiNoia2014}) and two types of novelty metrics (temporal novelty considering the timespan from the last object's update and user-perceived novelty describing the fraction of recommended objects, which were previously visited by the user).

However, as certain types of algorithms may provide recommendations that lack sufficient novelty or diversity, we also utilized strategies enhancing temporal novelty and diversity. Both novelty and diversity enhancements were applied as a post-processing of the lists of recommended objects. For diversity enhancements, we adopted the Maximal Margin Relevance approach \cite{Carbonell:1998:UMD:290941.291025} with $\lambda$ parameter held constant at $0.8$ and item-to-item similarity defined as a cosine similarity of their CB attributes. For enhancing temporal novelty, we re-ranked the list of recommended objects based on a weighted average of their original relevance $r$ and temporal novelty $novelty_t$:
\begin{equation}
\label{eq:novelty_enhancements}
    \bar{r}_o = \lambda*r_o + (1-\lambda)*novelty_t(o)
\end{equation}
$Novelty_t$ applies a logarithmic penalty on the time passed from the last object's update (see Eq. \ref{eq:temporal_novelty}). The $\lambda$ parameter was held constant at $0.8$.

\medskip

As the choice of a recommending algorithm, user's history aggregation, novelty and diversity enhancements are orthogonal, we run the off-line evaluation for all possible combinations. In total, 800 variants of RS were evaluated. 

\section{Evaluation Scenario}
In this section, we would like to describe the evaluation scenario and metrics. We separate the evaluation into two distinct parts: off-line evaluation on historic data and on-line A/B testing on a production server. 

\subsection{Off-line Evaluation}
\label{sec:off_line_eval}
For the off-line experiment, we recorded users' visits for the period of two and half years. The dataset contained over 560K records from 370K users. However, after applying restrictions on the volume of visits\footnote{Only the users with at least 2 and no more than 150 visited objects were kept.}, the resulting dataset contained 260K records of 72K users. We split the dataset into a train set and a test set based on a fixed time-point, where the interactions collected during the last month and half were used as a test set. The test set was further restricted to only incorporate users, who have at least one record in the train set as well, resulting into 3400 records of 970 users. 

In evaluation, we focused on four types of metrics, commonly used in recommender system's evaluation: rating prediction, ranking prediction, novelty and diversity. We evaluated several metrics for each class. 

For rating prediction, we suppose that visited objects have the rating $r=1$ and all others $r=0$. Mean absolute error (MAE) and coefficient of determination ($R^2$) were evaluated. 

For ranking-based metrics, we supposed that the relevance of all visited objects is equal, $r=1$ and other objects are irrelevant, $r=0$. Following metrics were evaluated: area under ROC curve (AUC), mean average precision (MAP), mean reciprocal rank (MRR), precision and recall at top-5 and top-10 recommendations (p5, p10, r5, r10) and normalized discounted cumulative gain at top-10, top-100 and a full list of recommendations (nDCG10, nDCG100, nDCG). The choice of ranking metrics reflects the importance of the head of the recommended list (p5, p10, r5, r10, nDCG10, MRR, MAP) as only a short list of recommendations can be displayed to the user. However, as the list of recommendable objects may be restricted due to the current context of the user (e.g., currently browsed category), we also included metrics evaluating longer portions of the recommended lists (AUC, nDCG100, nDCG).

As discussed in section \ref{sec:novelty_enhancements}, we distinguish two types of novelty in recommendations: recommending recently created or updated objects (temporal novelty) and recommending objects not seen by the user in the past (user novelty). For temporal novelty, we utilized logarithmic penalty on the timespan between current date and the date of the object's last update:
\begin{equation}
\label{eq:temporal_novelty}
    novelty_t = 1/(log(timespan.days) + \epsilon)
\end{equation}
 Mean of $novelty_t$ for top-5 and top-10 recommendations was evaluated. For user novelty, a fraction of already known vs. all recommended objects was used: $novelty_u =1 - |o \in \textrm{top-k} \cap o \in \textrm{known}_u |/k$  and evaluated for top-5 and top-10 objects. Finally, the intra-list diversity (ILD) \cite{DiNoia2014} evaluated at top-5 and top-10 recommendations was utilized as a diversity metric. 

All off-line metrics were evaluated for each pair of user and recommender. Mean values for each recommender are reported.

\begin{figure*}
\includegraphics[width=0.9\textwidth]{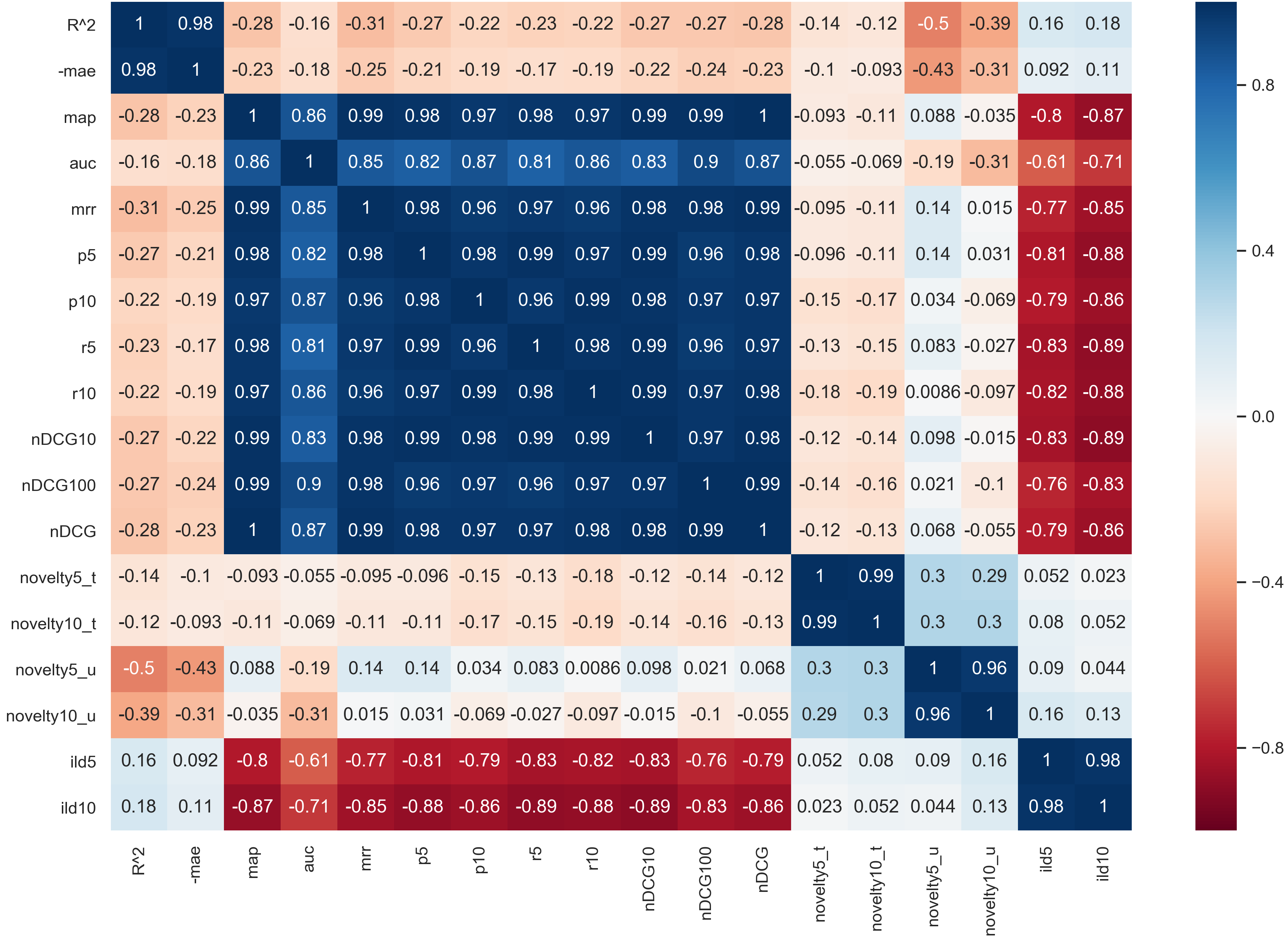}
\caption{Spearman's correlation for off-line evaluation metrics.}
\label{fig:results_offline_cor}
\end{figure*}

\subsection{On-line Evaluation}
The on-line A/B testing was conducted on the travel agency's production server during the period of one month. Out of 800 RS variants evaluated off-line, we selected in total 12 recommenders with (close to) best and (close to) worst results w.r.t. each evaluated metric. Details of the selection procedure are in section \ref{sec:offline_results}. One recommender was assigned to each user, based on his/her ID (i.e. $UID \% 12$). During the on-line evaluation, we monitored which objects were recommended to the user, whether (s)he clicked on some of them and which objects (s)he visited. The website tracks individual users\footnote{To be more specific, the website tracks a combination of a computer and a browser.} via cookies and does not require any registration or sign in in order to browse the tours. Therefore, we do not have any additional information about the users beyond their implicit feedback.

Based on the collected data, we evaluated two metrics: click through rate (CTR) and visit after recommend rate (VRR). CTR is a fraction between the volume of clicked and recommended objects and indicates that a recommendation was both relevant for the user and successful in catching his/her attention. 
VRR is a weaker criterion capturing situations, where after an object was recommended to the user, (s)he eventually visited it later on. 
In VRR, users might not saw recommendations, they might not fit his/her current context or the presentation was not so persuasive, however the recommended objects themselves were probably relevant). Although VRR is generally weaker than CTR, we utilized it for two primary reasons. The volume of collected feedback is considerably higher for VRR and as recommended objects were often placed outside of the initially visible area, CTR results may underestimate the true utility of recommendations. Note that if the object was recommended multiple times before the visit, we attribute the visit to the last recommendation.

\section{Results and Discussion}
\subsection{Off-line Evaluation}
\label{sec:offline_results}
Our aim in off-line evaluation was threefold. First, determine whether all evaluated metrics are necessary and provide valuable additional information. Second, identify, whether there are some general trends on the sub-classes of evaluated approaches or consistently dominating recommenders and finally, select suitable candidates for on-line evaluation.

We constructed matrices of Pearson's and Spearman's correlations for all considered off-line metrics. As both matrices are highly similar, we only report Spearman's correlation (see Figure \ref{fig:results_offline_cor}) to save space. The figure reveals several interesting patterns. Both diversity and rating prediction metrics are anti-correlated with ranking prediction metrics. The relation is especially strong for diversity. Novelty metrics are orthogonal to ranking accuracy as well as diversity and anti-correlated with rating prediction metrics. These results are somewhat similar to \cite{RS2015_news}, but individual clusters of metrics were less correlated in our case. We found the results of ranking vs. rating-based metrics quite consistent with findings of Herlocker et al. \cite{Herlocker2004_evaluations}.  Metrics from rating prediction, temporal novelty, user novelty and diversity classes were highly correlated ($\rho \geq 0.96$) and therefore only one metric for each category was selected (MAE, $\textrm{novelty10}_t$, $\textrm{novelty10}_u$, ILD10). As for ranking-based metrics, results were slightly more diversified. AUC was less correlated with all other ranking accuracy metrics ($0.81 \geq \rho \geq 0.9$), while for all other metrics $\rho \geq 0.96$. Pearson's correlation further separated $\{$nDCG100, nDCG$\}$ from the cluster of remaining ranking-based metrics and render them closer to the AUC. Therefore, we consider three clusters of ranking accuracy metrics: $\{$AUC$\}$, $\{$MAP, MRR, p5, p10, r5, r10, nDCG10$\}$ and $\{$nDCG100, nDCG$\}$. AUC, MRR and nDCG100 metrics were selected as representatives of each cluster.\footnote{Note that in order to illustrate differences among metrics, we occasionally display some additional ranking-based metrics in results.}

We further evaluated metrics correlations for individual recommending algorithms separately. Although the results were similar in general, there were some notable differences. $\textrm{Novelty}_u$ positively correlated with ranking accuracy metrics if evaluated for each recommending algorithm separately. The relation is strong especially for cosine CB recommenders.
 We also observed positive correlation between AUC and diversity as well as diversity and temporal novelty for cosine CB and correlation of temporal novelty with AUC for doc2vec. On the other hand, negative correlation between ranking-based metrics and both rating-based and diversity metrics was particularly strong for word2vec. Based on these observations, it may seem tempting to predict on-line performance for each algorithm separately, however the cold-start problem arises every time a new recommending algorithm has to be evaluated. Therefore, we did not follow this option and aimed on general prediction models based solely on off-line evaluation results.

Next, we evaluated individual recommender results according to the restricted set of metrics. First thing to note is that results were quite diverse. If a common ordering of metrics' results is considered (e.g., less MAE is better) 547 out of 800 recommenders were on the Pareto front. Therefore, we focused on providing some insight on recommending algorithms. Table \ref{tab:offline_results_recs} contains mean results as well as results of the best and worst member for each type of recommending algorithm. We may observe that while doc2vec models were superior in ILD, word2vec and cosine similarity performed considerably better w.r.t. ranking-based metrics. Furthermore, ILD score of doc2vec and word2vec were more than double than cosine similarity ones in average. 

As for the history aggregation methods, we observed that shorter history profiles provides considerably higher user-perceived novelty score. On the other hand \textit{max} history aggregations provided lowest $\textrm{novelty10}_u$ scores in average. We also observed that slightly better results w.r.t. ranking-based metrics achieved recommenders utilizing major portion of user's history (mean, temporal, temporal-10, last-10). Furthermore, recommenders with temporal-based user profiling also exhibited higher values of $\textrm{novelty10}_t$. Both diversity and novelty enhancements considerably increased ILD and $\textrm{novelty10}_t$ respectively with a negligible impact on other metrics. In general, type of the algorithm (cosine, word2vec, doc2vec) seems to have a determining impact on ranking-based and diversity metrics, surpassing effects of history aggregation, novelty enhancements or diversity enhancements.

\begin{table*}[t]
\caption{Off-line results for recommending algorithm types. {\normalfont Mean / Max / Min scores are depicted.}}
 \label{tab:offline_results_recs}
\begin{tabular}{llllllll}
 \toprule
Algorithm   & MAE               & AUC       & MRR      & nDCG100  & $\textrm{novelty10}_t$  & $\textrm{novelty10}_u$  & ILD10  \\
\midrule
doc2vec            & 0.37 / 0.46 / 0.21 & 0.58 / 0.72 / 0.52 & 0.02 / 0.06 / 0.01 & 0.05 / 0.10 / 0.03   & 0.23 / 0.30 / 0.21    & 0.79 / 0.91 / 0.57 & 0.80 / 0.89 / 0.58 \\
cosine			   & 0.40 / 0.42 / 0.36 & 0.78 / 0.80 / 0.74 & 0.14 / 0.19 / 0.07 & 0.21 / 0.24 / 0.17    & 0.23 / 0.27 / 0.22    & 0.87 / 0.97 / 0.57 & 0.26 / 0.44 / 0.20 \\
word2vec           & 0.36 / 0.42 / 0.22 & 0.81 / 0.85 / 0.73 & 0.09 / 0.15 / 0.04 & 0.19 / 0.25 / 0.11    & 0.23 / 0.29 / 0.21    & 0.74 / 0.89 / 0.57 & 0.59 / 0.85 / 0.42 \\
\bottomrule	
\end{tabular}
\end{table*}
		
While selecting candidates for on-line A/B testing, our main task was to determine predictability of on-line results from off-line metrics. However, due to the limited time and available traffic, the volume of recommenders evaluated in on-line A/B testing cannot be too high. 

Therefore, we adopted a following strategy: for each off-line metric, we selected the best and the worst performing recommender by default. However, if another recommender achieved close-to-best / close-to-worst performance
 and was already present in the set of candidates, we selected this one to save space. Furthermore, if a different type of algorithm achieved close-to-best performance, we considered its inclusion as well for the sake of diversity. Table \ref{tab:online_candidates} contains the final list of candidates for on-line evaluation.

\begin{table*}[t]
\caption{On-line and off-line results of recommenders selected for A/B testing. {\normalfont Div. and Nov: stands for diversity and novelty enhancements; parameter $e$ stands for embeddings size, $w$ denotes context window size and $s$ denotes whether calculating similarity on self is allowed. Best results w.r.t. each metric are in bold. For on-line metrics, results for users with 1-5 previously visited objects are depicted.}}
 \label{tab:online_candidates}
\begin{tabular}{lllll|lllllll|ll}
 \toprule

Algorithm   & Parameters               & History       & Nov.     & Div. & MAE & AUC       & MRR      & nDCG100  & $\textrm{nov10}_t$ & $\textrm{nov10}_u$ & ild10    & CTR      & VRR           \\
\midrule
\ \ 1: doc2vec      & e: 128, w: 1        & last     & yes & no  & 0.29 & 0.62 & 0.03 & 0.06    & 0.24     & 0.91 & 0.80 & 0.0071 & 0.0171 \\
\ \ 2: doc2vec      & e: 128, w: 1       & temp. & no  & yes & 0.36 & 0.68 & 0.03 & 0.08     & 0.22     & 0.74 & 0.83 & 0.0079 & 0.0200 \\
\ \ 3: doc2vec      & e: 32, w: 5         & mean     & no  & no  & 0.46 & 0.55 & 0.03 & 0.05     & 0.21     & 0.82 & 0.79 & 0.0089 & 0.0179 \\
\ \ 4: doc2vec      & e: 32, w: 5         & mean     & no  & yes  & 0.46 & 0.55 & 0.03 & 0.05     & 0.22     & 0.84 & \textbf{0.86} & 0.0063 & 0.0151 \\
\ \ 5: doc2vec      & e: 128, w: 5        & max      & yes & no   & \textbf{0.21}& 0.53 & 0.01 & 0.03    & 0.23     & 0.57 & 0.74 & 0.0073 & 0.0179 \\
\ \ 6: cosine 		& s:False 			& temp. & yes & no  & 0.40 & 0.80 & 0.14 & 0.21     & \textbf{0.26}     & \textbf{0.96}& 0.28 & 0.0056 & 0.0092\\
\ \ 7: cosine 		& s:True   			& mean     & yes & no & 0.40 & 0.80 & \textbf{0.15} & 0.21     & 0.23     & 0.80 & 0.22 & \textbf{0.0112} & \textbf{0.0218} \\
\ \ 8: cosine 		& s:True   			& last-10 & no  & no  & 0.39 & 0.78 & 0.13 & 0.20    & 0.22     & 0.80 & 0.21 & 0.0073 & 0.0166 \\
\ \ 9: word2vec    & e: 64, w: 5          & mean     & no  & yes  & 0.37 & 0.83 & 0.11 & 0.20     & 0.22     & 0.72 & 0.67 & 0.0095 & 0.0206 \\
10: word2vec    & e: 32, w: 5        & temp.     & no  & yes   & 0.42 & \textbf{0.84} & 0.14 & 0.22     & 0.25     & 0.78 & 0.48 & 0.0095 & 0.0198 \\
11: word2vec    & e: 128, w: 3        & last     & no  & no  & 0.29 & 0.75 & 0.10 & 0.17     & 0.22     & 0.85 & 0.51 & 0.0068 & 0.0173 \\
12: word2vec    & e: 32, w: 3         & last-10 & no  & no   & 0.42 & \textbf{0.84} & 0.12 & \textbf{0.23}     & 0.22     & 0.75 & 0.42 & 0.0082 & 0.0186 \\
\bottomrule	
\end{tabular}
\end{table*}

\subsection{On-line Evaluation}
\begin{figure}
\includegraphics[width=8.5cm]{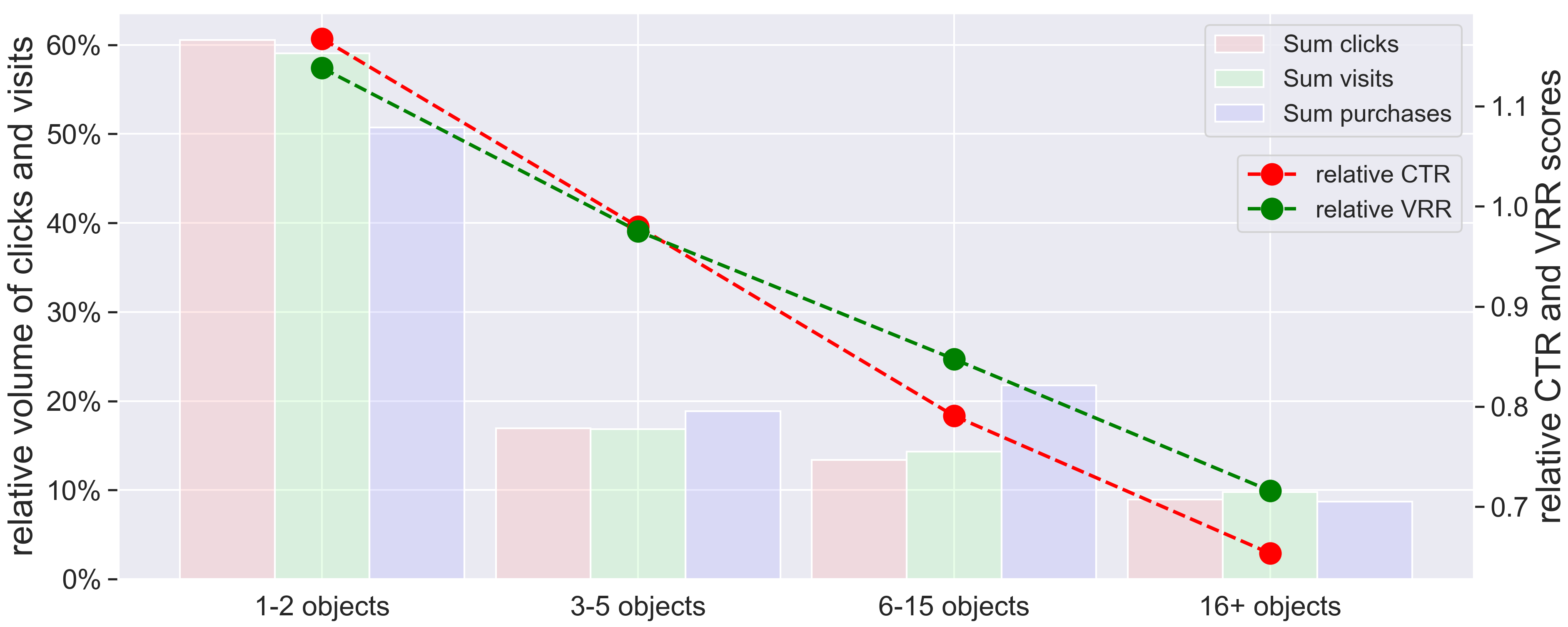}
\caption{Comparing relative volumes of visits, clicks and purchases as well as relative CTR and VRR for different user profile sizes w.r.t. the overall values. }
\label{fig:relative_ctr_vrr}
\end{figure} 

A total of 4287 users participated in the on-line evaluation, to whom, a total of 130261 objects were recommended\footnote{We excluded global-only recommendations provided to users without any past visited objects and results of users with too many visited objects (probably agency's employees).}. The total volume of click-through events was 928 and the total volume of visits after recommendation was 2102. The difference between the volume of clicks and visits illustrates the problem of recommendations discoverability or ability to catch user's attention. As these features may be partially deduced from the implicit feedback data \cite{rev-PV-JDataSem}, we plan to incorporate off-line models that consider objects' discoverability in the future work. Another possibility is that recommended objects were potentially relevant, but not in the current context (user eventually process them after some time). Also this factor may be revealed by a more detailed implicit feedback analysis in the future. 

While processing the results, we observed that they are strongly conditioned by the "seniority" of users measured as the volume of previously visited objects. This is illustrated on Figure \ref{fig:relative_ctr_vrr}, where four sets of users with \textit{1 - 2}, \textit{3 - 5}, \textit{6 - 15} and \textit{16+} previously visited objects are distinguished.
The highest overall volume of interactions (clicks, visits, purchases) was collected for novice users with 1-2 visited objects. This group also exhibits highest CTR and VRR rates if compared with average values. Relative CTR and VRR drops for users with larger profiles (CTR exhibits slightly steeper decrease). On the other hand, we may see that purchase volumes did not decrease as much as other types of interactions for users with 3-15 visited objects, which shows importance of these "moderately senior" group of users from the business perspective.


\begin{figure*}
\includegraphics[width=14cm]{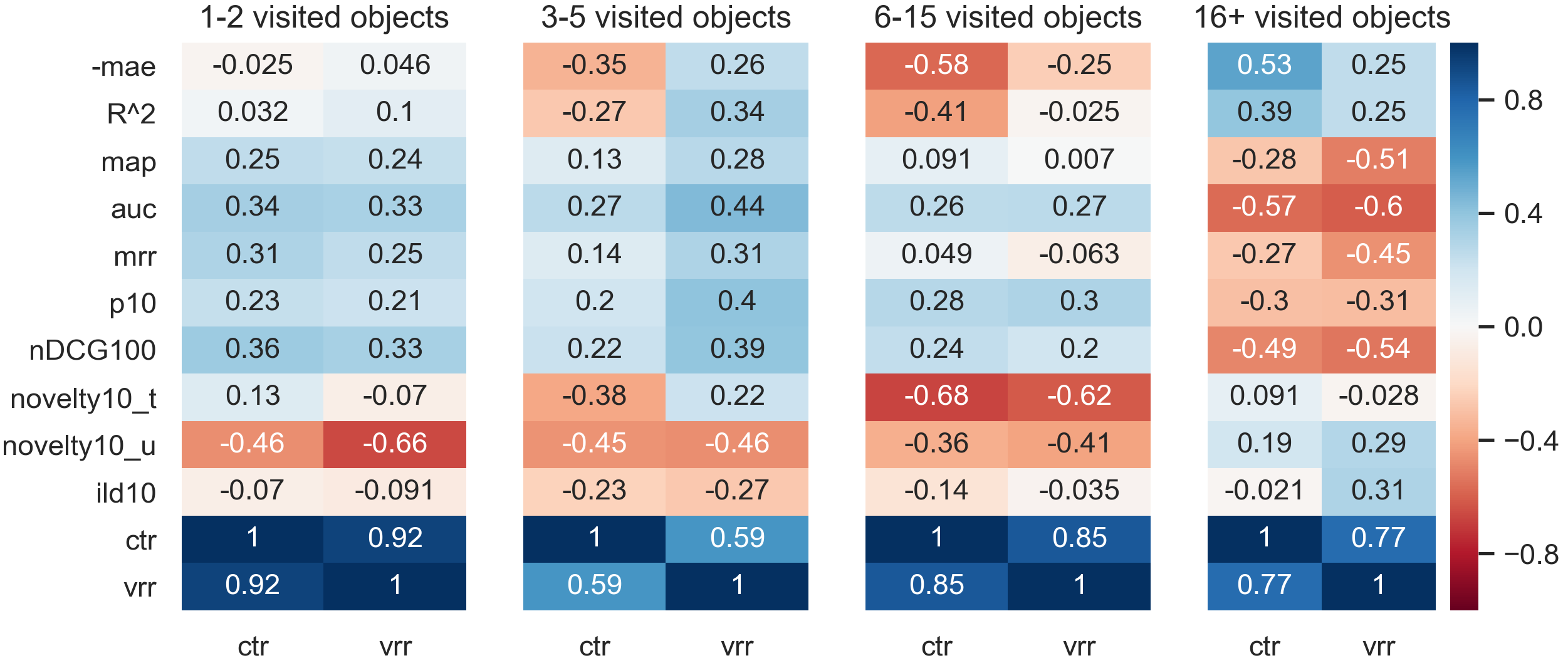}
\caption{Spearman's correlation between off-line and on-line evaluation metrics for various user model sizes.}
\label{fig:results_online_cor}
\end{figure*}

Table \ref{tab:online_candidates} contains results of on-line A/B testing (VRR and CTR) for individual recommending algorithms and users with 1 - 5 visited objects. In general, doc2vec variants performed slightly worse than word2vec w.r.t. both CTR and VRR. Both overall best and worst algorithm belongs to the Cosine CB family. We suppose that the exceptionally bad performance of algorithm ID 6 was caused by too high user-perceived novelty (caused by $s:False$ hyperparameter). There are some related works with similar conclusions, e.g. Herlocker et al. \cite{Herlocker2004_evaluations} suggested that users may require a certain portion of known items to be present in the recommendations in order to trust the recommender. Also Jannach et al. \cite{Jannach2017SessionbasedIR} observed that reminders (i.e. known items) exhibit higher CTR than other forms of recommendations. Nonetheless, we plan to verify this hypothesis in the future work.

Figure \ref{fig:results_online_cor} depicts Spearman's correlation between on-line and off-line evaluation metrics for users with \textit{1 - 2}, \textit{3 - 5}, \textit{6 - 15} and \textit{16+} visited objects. 
We may observe a significant twist in the performance according to the seniority of users. While for novel users, ranking-based metrics exhibits some correlation to both on-line metrics, this starts to decrease for more senior users (6-15 objects) and finally turns into a negative correlation for users with 16+ visited objects. An opposite behavior can be seen for user-perceived novelty and partially also for ILD. 

We hypothesize that more senior users might already observed most of the straightforward choices (the evaluation site contained rather low volume of objects and provides a broad palette of browsing options). Therefore, novel and diverse suggestions may be appropriate for them. Again, we plan to focus on this hypothesis in the future work. Similarly, we plan to further investigate the rather surprising behavior of rating-based metrics as no clear pattern can be seen at the moment.

\begin{figure}
\includegraphics[width=8.5cm]{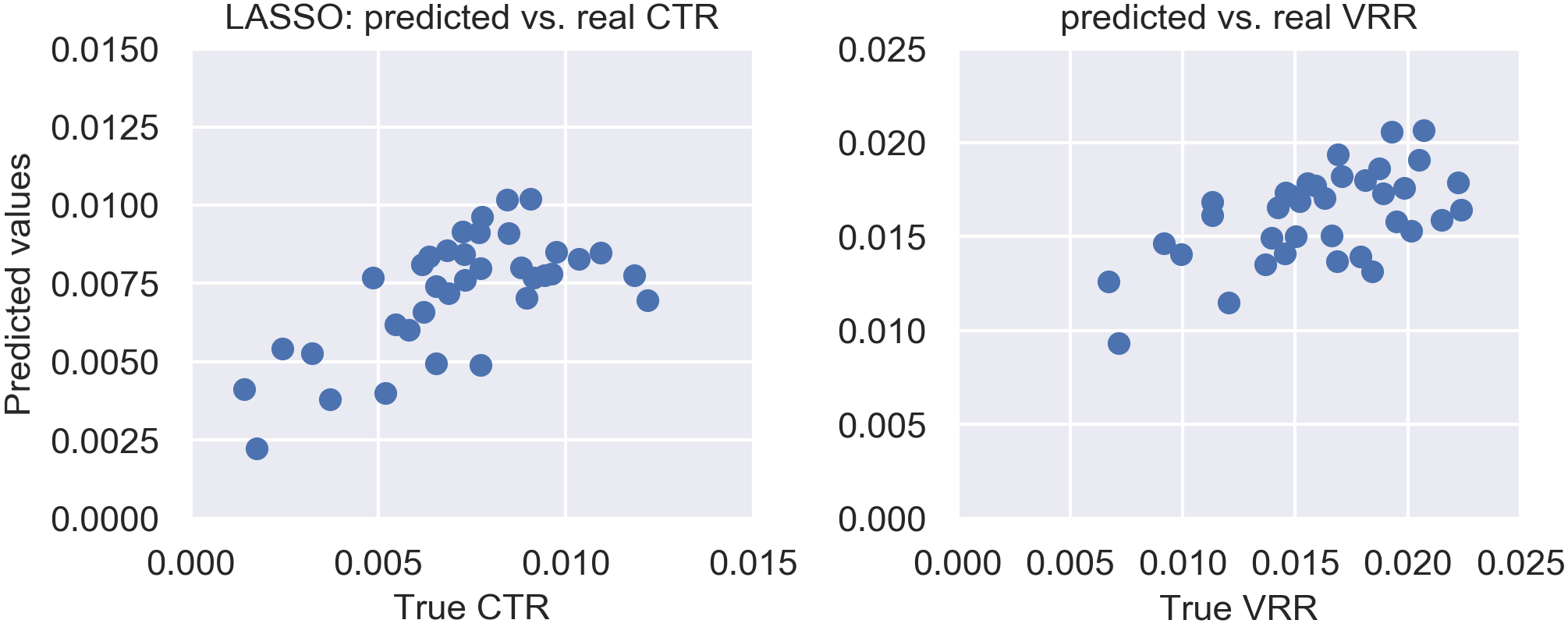}
\caption{True values of CTR and VRR compared with the ones predicted via LASSO regression. }
\label{fig:predicted_vs_true_ctr}
\end{figure} 

\subsection{Results Post-processing}

After the completion of on-line experiments, we also aimed to revisit previous off-line results with the knowledge from on-line / off-line comparison. In order to do so, we trained several simple regression methods aiming to predict CTR and VRR from off-line evaluation metrics. Because of the twist in on-line performance for increasing user profile sizes, we decided to treat the on-line results separately for for users with \textit{1 - 2}, \textit{3 - 5} and \textit{6 - 15} visited objects\footnote{We were unable to learn any reliable predictor for the group ot users with 16+ objects, therefore we exclude it from the results.}. We further incorporate the user profile size into the set of input variables. Due to the very small dataset size (12 algorithms $\times$ 3 user profile size groups = 36 data points), we only focused on simple regression techniques to prevent over-fitting. We evaluated linear regression, LASSO and decision tree predictions according to the leave-one-out cross validation (LOOCV) scheme and degree-2 polynomial input feature combinations. 

Due to very high coefficients, linear regression often predicted unrealistic values for both CTR and VRR (i.e. CTR $\ll$ 0 and $\gg$ 1). Decision tree often failed to provide reasonable predictions and (unsurprisingly) constructs large sets of algorithms with equal predicted values. However, with LASSO regression model, we were able to predict both on-line metrics up to some extent (see Figure \ref{fig:predicted_vs_true_ctr}. Specifically, $R^2$ scores were 0.42 and 0.35 for CTR and VRR respectively, while Kendal's Tau-b scores were 0.39 and 0.4 (in both cases, p-value < 0.05). 

We also evaluated prediction of LASSO for the original set of 800 recommending algorithms. Among the top-20 results, word2vec models and cosine CB models were present, often with \textit{max} history aggregations or some variant of \textit{temporal} history aggregations. 

Finally, in our last experiment, we aimed on verifying the quality of the CTR and VRR prediction models. Therefore, we run one more iteration of the on-line A/B testing. The best-performing model from the previous phase (ID 7 from Table \ref{tab:online_candidates}) served as a baseline in this test. Furthermore, we included two variants (for CTR and VRR) of the best algorithms according to LASSO regression (the individual algorithm was selected according to the actual user profile size). 

Results of this experiment were unfortunately rather inconclusive. The original Cosine CB model scored 0.0064 and 0.0167 for CTR and VRR respectively. LASSO-predicted algorithms scored 0.0069 for CTR and 0.0185 for VRR. However, in both cases, the differences were not statistically significant. Nonetheless, we may conclude, that the prediction methods managed to provide candidates comparable with the so-far best method (we hope to provide more conclusive results for the camera-ready).

\section{Conclusions and Future Work}
In this paper, we conducted an extensive comparison of off-line and on-line evaluation metrics in the context of small e-commerce enterprises. Experiments were held on a Czech 
medium-sized travel agency and shown a moderate correlation between ranking-based off-line metrics (AUC, MRR, P10, nDCG100) and both visits after recommend rate (VRR) and click-through rate (CTR) for less senior users. Similarly, results indicated a negative correlation between on-line metrics (CTR, VRR) and user-perceived novelty for the same group of users. Nonetheless, these results are reversed for the more senior users, which may indicate their saturation with simple suggestions. 

However, further work is needed to verify, whether this relation may be caused by the choice of recommending algorithms, or whether there are user or object clusters with different behavior.

In addition to the direct on-line - off-line comparison, we trained several regression models aiming to predict on-line results from off-line metrics. Some of these models achieved reasonable performance for both CTR and VRR and we were able to select good additional candidate recommenders for A/B testing. 

Our future work should include more detailed analysis of algorithms' off-line performance w.r.t. different segments of users and also incorporating relevant contextual information into the evaluation process. Furthermore, we plan to evaluate additional hybrid or ensemble approaches utilizing multiple sources of information (CB, CF) as well as session-based recommendations. An interesting point to observe is, to what extent the on-line results can be predicted also for some new classes of recommending algorithms.

Our future work should also incorporate utilization of more complex implicit user feedback in order to assess importance of visited objects as well as decrease the visibility noise in on-line evaluation, especially CTR. Finally, in order to provide more transferable knowledge, we plan to perform similar experiments also on some additional small e-commerce enterprises.

\begin{acks}
This paper has been supported by Czech Science Foundation (GA\v{C}R) project Nr. 19-22071Y and by Charles University project Progres Q48. Source codes, evaluation data and complete results are available from \url{github.com/lpeska/HT2020}.
\end{acks}

\bibliographystyle{ACM-Reference-Format}
\bibliography{references}

\end{document}